# Maxwell's fisheye lens as efficient power coupler between dissimilar photonic crystal waveguides

S. Hadi Badri[a,]*, M. M. Gilarlue[a]

[a] Department of Electrical Engineering, Sarab Branch, Islamic Azad University, Sarab, Iran

Corresponding author.

* E-mail addresses: sh.badri@iaut.ac.ir

## Abstract

The imaging properties of the Maxwell's fisheye (MFE) lens makes it a viable candidate to implement power coupling between different types of waveguides. A coupler based on the MFE lens is designed to couple a square lattice photonic crystal to a triangular lattice one. The MFE lens is implemented as ring-based multilayer and graded photonic crystal (GPC) structures. The performance of the ring-based MFE lens is better than the GPC-based one in the C-band of optical communication. The proposed ring-based MFE coupler has a footprint of $3.62 \mu m \times 3.62 \mu m$ and covers the entire C and U bands. The S and L bands are partially covered. The average insertion loss of $0.1 dB$ and the maximum return loss of $-11 dB$ in the C-band is achieved.

## Keywords

## 1. Introduction

Photonic crystals (PhC) are an interesting branch in photonic integrated circuits (PIC). There are different types of PhC lattices which have their own merits and applications [1-4]. Coupling an optical signal from one type of lattice to another type of lattice may be required in designing of complex PICs [5, 6]. The modal mismatch between these different types of PhC lattices decreases coupling efficiency. There have been variety of methods proposed to couple a channel waveguide to PhC waveguide such as butt-coupling [7, 8], mode-field matching [9-11], anti-reflection layer [12], semi-adiabatic coupler [13], topology optimization [14, 15]. However, few methods have been introduced to couple different PhC waveguides to each other such as inverse optimization [5] and resonant cavities [6] methods. The method in [5] is based on grating-like surfaces that are optimized with generic algorithms coupling square and triangular lattice PhC waveguides. The bandwidth of 20nm centered at 1552nm, minimum insertion loss of $1 dB$, and $4.8 \mu m \times 9.0 \mu m$ footprint has been reported with this method. In [6] standard linear defect, triple-wide linear defect, and coupled-cavity waveguides are coupled together based on the resonant cavities with a footprint of few lattice constants. Resonant cavities inherently have limited bandwidth. Therefore, limited bandwidth of this method restricts its application to other waveguide types.

The graded-index (GRIN) lenses such as the MFE lens have been used for designing of different optical components [16]. Image of a point source on the surface of the MFE lens is reproduced on the diametrically opposite side of the lens. Recently, the MFE lens is used to design single-mode and multimode waveguide crossings [17, 18]. Due to the imaging properties of the MFE lens, it could also be used to design a power coupler. In this paper, a power coupler based on the MFE lens is presented that couples the optical wave from PhC of square lattice to triangular one. The refractive index of the MFE lens is

$$n_{lens}(r) = \frac{2 \times n_{min}}{1+(r/R_{lens})^2} \quad , \quad (0 \leq r \leq R_{lens}) \tag{1}$$

where $R_{lens}$ is the radius of the lens and $r$ is the radial distance from the center of the lens, and $n_{min}$ is the refractive index of the lens at its edge.

## 2. Implementation of the MFE lens

The effective medium theory can be used to design a medium with a range of refractive-indices between the refractive indices of the constituent materials. When the long-wavelength limit is met the subwavelength structure can be considered as a homogeneous medium. The effective refractive index of the composite material depends on the electric field direction with respect to the arrangement of the inclusion layers in the host. When the inclusions layers are parallel to the electric field, the effective refractive index is approximated by [19]

$$n_{eff,TM}^2 = f_{inc} n_{inc}^2 + (1 - f_{inc}) n_{host}^2 \tag{2}$$

where $n_{host}$, $n_{inc}$, and $n_{eff,TM}$ are the refractive indices of the host, inclusion, and effective medium for TM mode, respectively. The filling factor, $f_{inc}$, is the fraction of the total volume occupied by inclusion layer. The dependence of the effective refractive index on the polarization of optical signal is called form birefringence [20]. It is possible to control the effective refractive index of the composite material by changing the filling factor of the inclusion layer. The presented concentric ring-based multilayer MFE lens is composed of SiO$_2$ and Si layers. In our design, SiO$_2$ and Si are considered as the host and inclusion layers, respectively. The procedure to design the lens is to divide it into concentric cylindrical layers of equal width, $dr$. To calculate the width of inclusion layer, $dr_{inc}$, in the i-th layer Eq. 2 is rearranged as

$$f_{inc} = \frac{n_{eff,TM}^2 - n_{host}^2}{n_{inc}^2 - n_{host}^2} \tag{3}$$

The filling factor for the i-th layer is $f_{inc} = A_{inc}/A_i$ where $A_{inc} = 2\pi r_i dr_{inc}$ and $A_i = 2\pi r_i dr$ are areas of inclusion and the i-th layers, respectively. The width of the inclusion layer is given by [17]

$$dr_{inc} = \frac{n_{eff,TM}^2 - n_{host}^2}{n_{inc}^2 - n_{host}^2} dr \tag{4}$$

To implement the lens with graded photonic crystal (GPC) a similar approach as described in above is followed. Each layer in the lens is divided into annular cells where a rod as an inclusion material is placed at the center of the cell. The radius of the rod in the ij-th cell is calculated by [18]

$$r_{rod,ij} = \sqrt{\frac{A_{ij}(\varepsilon_{eff,ij}^{TM} - \varepsilon_{host})}{\pi(\varepsilon_{inc} - \varepsilon_{host})}} \tag{5}$$

where $A_{ij}$ is the area of the *ij*-th cell, $\varepsilon_{eff}^{TM}$ is the effective permittivity of the cell for TM mode, $\varepsilon_{inc}$ and $\varepsilon_{host}$ are the permittivities of the inclusion and host materials, respectively.

## 3. Numerical simulation and discussion

The two-dimensional (2D) simulations were carried out with Comsol Multiphysics™ to investigate the MFE lens's performance as a power coupler. We want to interconnect two different kinds of PhC to each other. As shown in Fig.1 the square lattice PhC and triangular lattice PhC are coupled through the MFE lens. Both PhC structures consist of cylindrical *Si* rods embedded in *SiO₂* background. In our simulations, the refractive indices of *SiO₂* and *Si* are considered as *1.45* and *3.45*, respectively. The lattice constant of $a = 465nm$ and rods with a radius of $r = 0.2a$ are considered for both of the triangular and square lattices. Both PhC structures have a bandgap in TM mode covering S, C, L, and U optical communication bands. Since the width of waveguide in square lattice is larger than the triangular one, we have removed two rows in triangular lattice similar to [5]. The MFE lens's refractive index profile is implemented with concentric cylindrical layers and GPC as described in the previous section. To minimize the reflection from the interface of waveguides and the MFE lens, the lens's refractive index changes radially from *2.9* at the center to *1.45* at the edge of the lens to match the refractive index of the SiO₂ material. The radius of the MFE lens is $R_{lens} = 1.81 \mu m$, and it is divided into *9* cylindrical layers. The decrease in the width of inclusion layers (rods) from the center towards the edge of the lens corresponds to the decrease in refractive index of the MFE lens implemented with ring-based (GPC-based) structure. The concentric ring-based structure of the MFE lens is also shown in Fig. 1. The inclusion (Si) and the host (SiO₂) materials are specified in this figure. We have used perfectly matched layer (PML) domains behind the input and output ports to reduce the spurious reflection from the ports. This PhC-based PML domains have been described and illustrated in [17, 18] so they are not shown in Fig. 1. The out-of-plane component of electric field intensity is illustrated in Fig. 2 for the concentric ring-based structure at the wavelength of 1550*nm*. As seen in this figure, the phase front expands to a wide plane at the center of the MFE lens and again converges at the diametrically opposite point which is the entrance of the waveguide in the triangular lattice. At this wavelength, the insertion loss was 0.13dB and the return loss was -15.2dB.

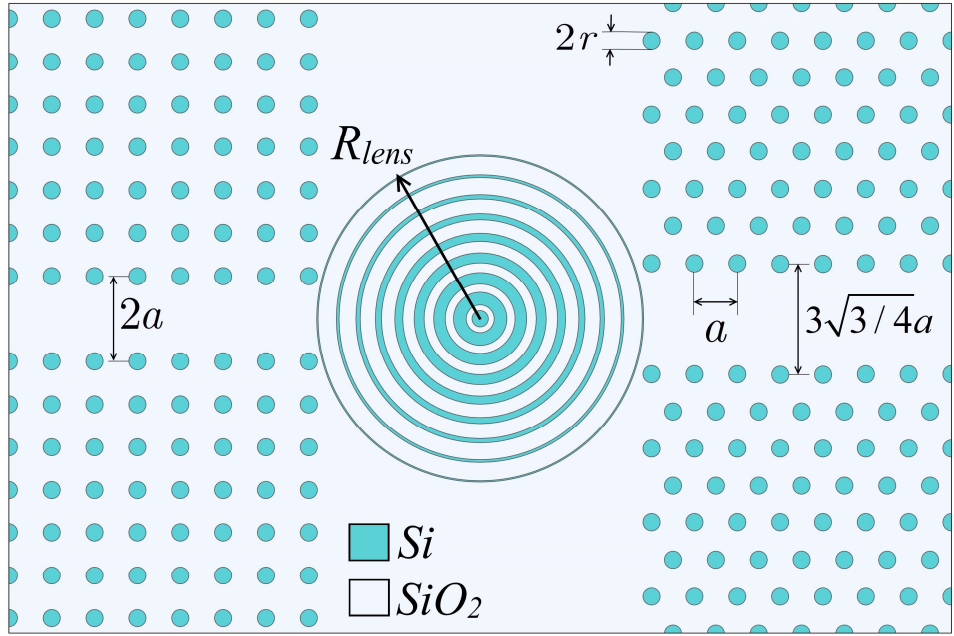

Fig. 1. Overview of geometrical structure of square and triangular lattices and the proposed ring-based MFE lens as power coupler

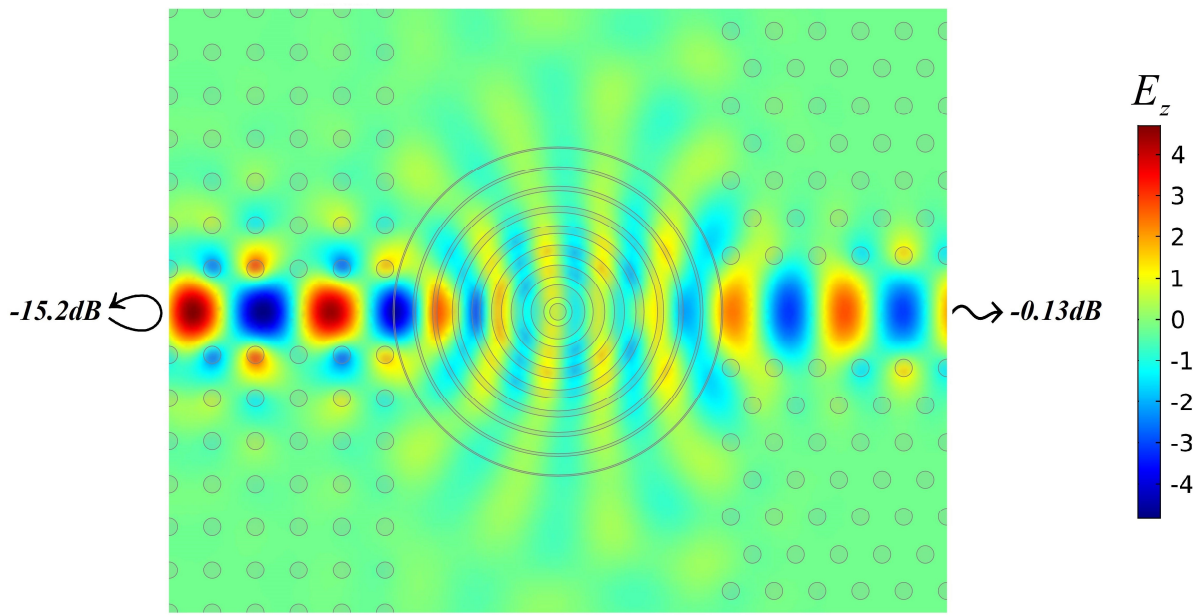

Fig. 2. Out-of-plane component of electric field at wavelength of 1550nm propagating through power coupler

Fig. 3 shows the GPC-based MFE lens as power coupler between two different types of PhCs. The power streamlines visualize the propagation of the average power flow from the square lattice waveguide to the triangular one. The GPC-based lens had the same radius and number of layers as the ring-based lens. The only difference between them was that the effective refractive index of each layer in the GPC-based lens was implemented by Si rods in the $SiO_2$ background. Coupling efficiency of both ring-based and GPC-based MFE lenses are compared through their scattering parameters in Fig. 4. The passbands of the ring-based lens are 1460-1489$nm$, 1496-1576$nm$, and 1586-1675$nm$. And the GPC-based lens covers the 1460-1481$nm$, 1484-1568$nm$, and 1586-1675$nm$ bands. For the ring-based lens the average insertion loss of 0.1$dB$ and the maximum return loss of -11$dB$ in the C-band was achieved. On the other hand, the average insertion loss of 0.8$dB$ and the maximum return loss of -5.7$dB$ in the C-band was obtained for the GPC-based lens. Although there are some differences in the performance of the two structures but both of the implementations cover the entire C and U bands while S and U bands are partially covered.

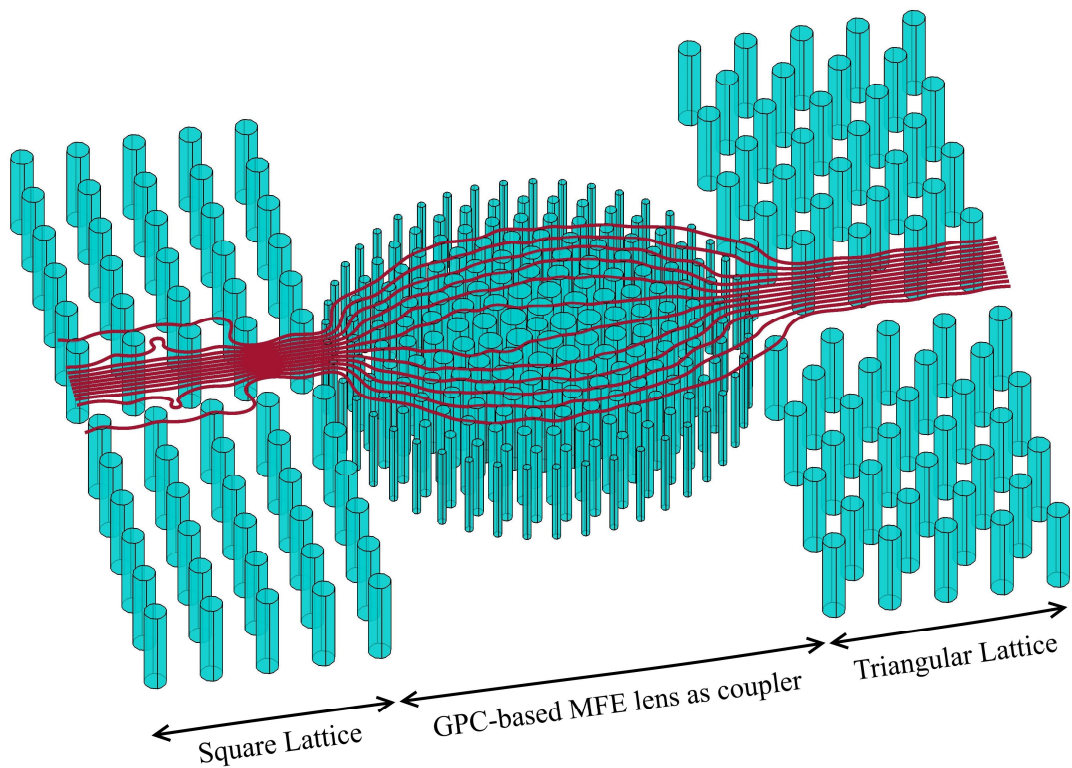

Fig. 3. Flow of power from square lattice to triangular lattice through the GPC-based MFE lens

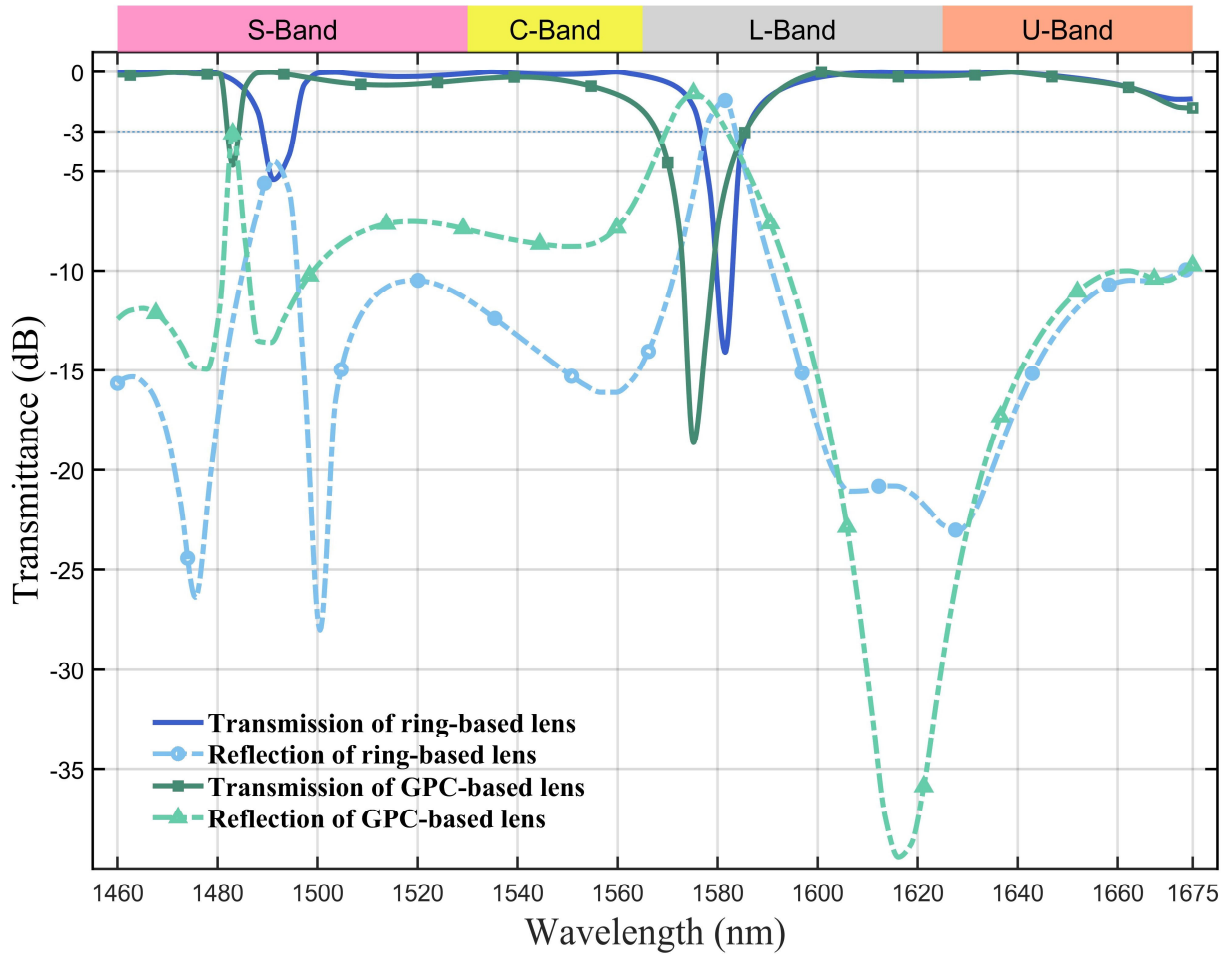

Fig. 4. Transmission and reflection of ring-based and GPC-based MFE lens as power coupler

In order to highlight the superior performance of the MFE lens as power coupler compared to previous methods we have regenerated the results of [5] and [6]. The performance of the ring-based MFE lens is compared with methods of [5] and [6] in Fig 5. The bandwidth of methods [5] and [6] are 20*nm* and 8*nm*, respectively. The insertion loss of the reference [5] was higher than 1.2*dB* while the reference [6] had insertion loss of 0*dB* at 1550nm. Our proposed method had a wider bandwidth and lower insertion loss.

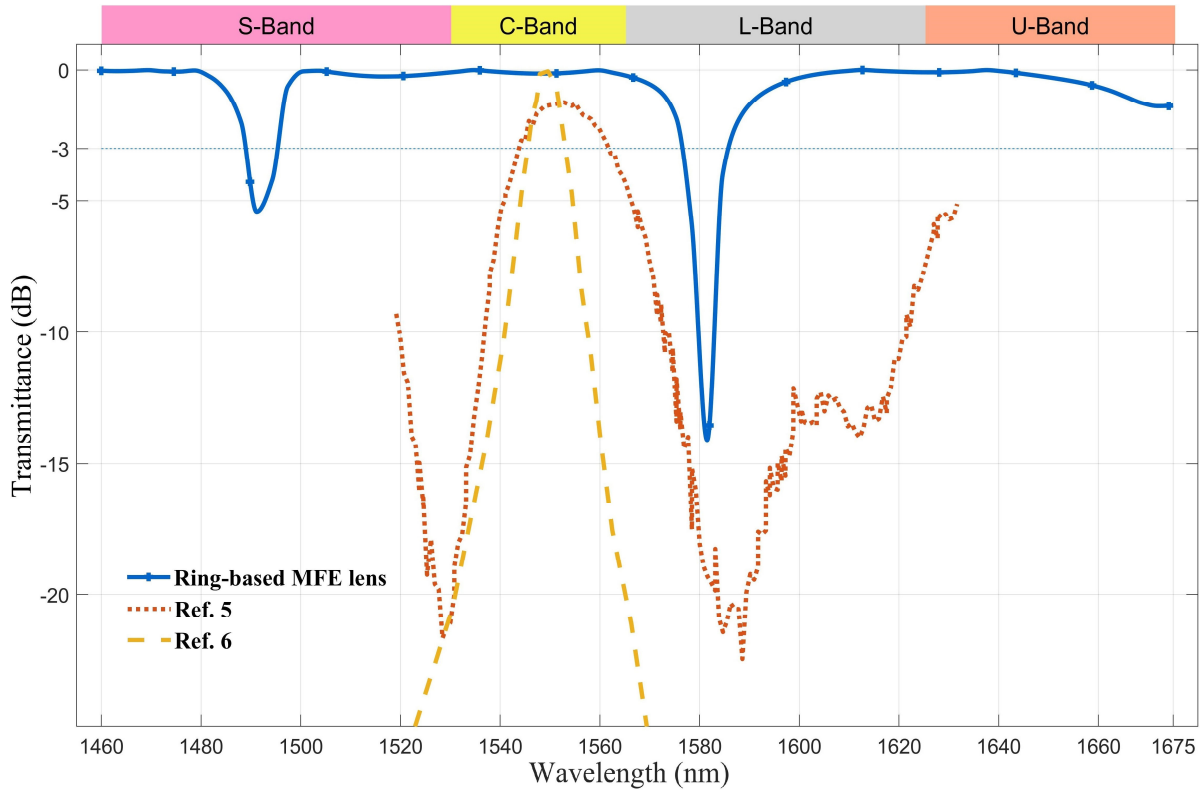

Fig. 5. Comparison of the ring-based MFE lens and references [5] and [6]

## 4. Conclusion

The square and triangular lattices are common types of PhCs used to design variety of optical devices. Each of the mentioned PhC lattices present their own merits. It may be required to couple the light signal from one type of PhC lattice to another one. The performance of the MFE lens as power coupler was investigated. The MFE lens was implemented with ring-based and GPC-based structures. For the ring-based (GPC-based) lens the average insertion loss of $0.1dB$ ($0.8dB$) and the maximum return loss of $-11dB$ ($-5.7dB$) in the C-band was achieved. The footprint of the proposed coupler was $3.62\mu m \times 3.62\mu m$.